\documentclass[aps,floatfix,superscriptaddress,twocolumn,pra]{revtex4}
\usepackage{graphicx,bm,color}
\usepackage{array}
\usepackage{amsmath,amsfonts,amssymb}
\usepackage{pdflscape}
\usepackage{hyperref}
\hypersetup{colorlinks=true}
\hypersetup{urlcolor=blue}
\hypersetup{citecolor=black}
\hypersetup{menucolor=black}
\hypersetup{linkcolor=black}
\usepackage{cleveref}
\usepackage{longtable}
\usepackage{float}
\usepackage{pdfpages}
\usepackage[T1]{fontenc}
\usepackage[utf8]{inputenc}
\usepackage{bm}
\usepackage{ulem}
\usepackage{tikz}
\usetikzlibrary{calc,patterns,decorations.pathmorphing,decorations.markings}

\newcommand{\be}{\begin{equation}}
\newcommand{\ee}{\end{equation}}
\newcommand{\bea}{\begin{eqnarray}}
\newcommand{\eea}{\end{eqnarray}}

\newcommand{\ket}[1]{|#1\rangle}
\newcommand{\meanv}[1]{\langle #1 \rangle}
\newcommand{\norm}{\mathcal{N}}

\newcommand{\bb}[1]{\left( #1 \right)}

\newcommand{\bbacol}[1]{\left\{ #1 \right.}
\newcommand{\bbacor}[1]{\left. #1 \right\}}
\newcommand{\bbcro}[1]{\left[ #1 \right]}

\newcommand{\ii}{\textrm{i}}
\newcommand{\dd}{\textrm{d}}
\newcommand{\eee}{\textrm{e}}
\newcommand{\rr}{\textbf{r}}
\newcommand{\qq}{\textbf{q}}
\newcommand{\kk}{\textbf{k}}
\newcommand{\zero}{\textbf{0}}
\newcommand{\kkin}{\textbf{k}\in\mathcal{D}}
\newcommand{\kkPin}{\textbf{k}'\in\mathcal{D}}

\newcommand{\PP}{\textbf{P}}

\newcommand{\resoumis}[1]{{#1}}

\tikzstyle{ressort}=[decorate,decoration={zigzag,pre length=0.0cm,post length=0.0cm,segment length=5, amplitude=0.1cm}]

\tikzset{->-/.style={decoration={
  markings,
  mark=at position .5 with {\arrow[scale=2,color=black]{>}}},postaction={decorate}}}
\tikzset{-<-/.style={decoration={
  markings,
  mark=at position .5 with {\arrow[scale=3,color=black]{<}}},postaction={decorate}}}

\tikzset{->>-/.style={decoration={
  markings,
  mark=at position .5 with {\arrow{>>}}},postaction={decorate}}}
\tikzset{-<<-/.style={decoration={
  markings,
  mark=at position .5 with {\arrow{<<}}},postaction={decorate}}}

\tikzset{phantom->-/.style={decoration={
  markings,
  mark=at position .5 with {\arrow[scale=2]{>}}},postaction={decorate}}}

\tikzset{serpent/.style={decoration={snake},postaction={decorate}}}

\begin{document}

\title{Absorption and emission of a collective excitation by a fermionic quasiparticle in a Fermi superfluid}
\author{Hadrien Kurkjian}
\affiliation{TQC, Universiteit Antwerpen, Universiteitsplein 1, B-2610 Antwerpen, Belgi\"e}
\author{Jacques Tempere}
\affiliation{TQC, Universiteit Antwerpen, Universiteitsplein 1, B-2610 Antwerpen, Belgi\"e}
\begin{abstract}
We study the process of absorption or emission of a bosonic collective excitation by a fermionic quasiparticle in a superfluid of paired fermions. From the RPA equation of motion of the bosonic excitation annihilation operator, we obtain an expression of the coupling amplitude of this process which is limited neither to resonant processes nor to the long wavelength limit. We confirm our result by independently deriving it in the functional integral approach using the gaussian fluctuation approximation, and by comparing it in the long wavelength limit to the quantum hydrodynamic result. Last, we give a straightforward application of the coupling amplitude we obtain by calculating the lifetime of the bosonic excitations of arbitrary wave number. We find a mode quality factor that decreases from its maximum at low wave numbers and vanishes when the bosonic branch hits the \resoumis{continuum of fermionic biexcitations}.
\end{abstract}
\maketitle


\section{Introduction}


At low temperature, when a quantum fluid is close to its ground state, it is generally described in terms of its elementary excitations or quasiparticles. These eigenmodes of the many-body system remain weakly populated as long as the gas is weakly excited. The quasiparticles are then weakly interacting, so that to first approximation one considers them as an ideal gas. Going beyond this description to include the coupling between quasiparticles is often a theoretical challenge but is essential to account for the ergodicity of the system and the dissipative phenomena such as the damping of the quasiparticles \cite{Beliaev1958}, the loss of macroscopic coherence \cite{CastinSinatra2009,CRAS2016} or the influence of temperature on the viscosity \cite{Khalatnikov1949} and the equation of state of the fluid \cite{Manuel2010}. At zero temperature, taking the interactions among quasiparticles into account to compute the quantum fluctuations in the elementary modes can also improve the description of the ground state of the fluid \cite{Castin2009}.

The quantum fluid we consider is a gas of ultracold spin-$1/2$ atomic fermions in which van der Waals interactions occur only in the $s$-wave channel between opposite spin fermions and favor the formation of $\uparrow/\downarrow$ pairs. Below the critical temperature, those pairs condense into a macroscopically occupied wave function called a pair condensate, and the gas becomes superfluid. This phenomenon is nowadays commonly observed in the laboratory  \cite{Thomas2002,Salomon2003,Grimm2004b,Ketterle2004}. The elementary excitations of this system are of two kinds : like in any superfluid, there exists a bosonic branch \cite{Anderson1958,CKS2006} that in the long wavelength limit describes sound waves \cite{Khalatnikov1949} whose energy is proportional to the wave number. Since the elementary constituents of the superfluid are pairs of fermions, there also exists a fermionic branch of excitation of the relative motion of the pairs, with a gapped spectrum \cite{BCS1957}.
Several important experimental results on the ground state and the elementary excitations of this system were obtained recently: the measurement of its zero temperature equation of state \cite{Salomon2010} and of the dispersion relation of the fermionic \cite{Ketterle2008} and bosonic \cite{Thomas2007,Stringari2013} excitations, also at high energy \cite{Vale2017}. Experiments can now probe rather precisely the physics of the gas at non-zero temperatures \cite{Zwierlein2012}, but its description with analytical methods remains a theoretical challenge. Among the existing approaches to determine the non-zero temperature equation of state we mention a phenomenological model \cite{Bulgac2006} based on the thermal occupation of the bosonic and fermionic branch, and, at unitarity, effective field theories exploiting the extra symmetries of the system \cite{SonWingate2006,Manuel2010}. The description of the interactions among quasiparticles relies on quantum hydrodynamics \cite{Khalatnikov1949}, a low-energy effective theory limited to the leading order in temperature, except for the coupling between three bosonic excitations \cite{Annalen}. In this framework, the resonant three- and four-body couplings between phonons \cite{Annalen,EPL} and between phonons and fermionic quasiparticles \cite{bosonsfermions} were obtained.

In this article, we compute the amplitude of the inelastic process of absorption or emission of a collective bosonic excitation by a fermionic quasiparticle within two distinct microscopic approaches leading to the same result : the Random Phase Approximation (RPA) \cite{Anderson1958} and the functional integral approach in the gaussian fluctuation approximation \cite{Drummond2006,Randeria2008}. Our result is in agreement with quantum hydrodynamics in its validity domain but is not limited to resonant processes nor to the long wavelength limit. We present a straightforward application: the computation of the damping rate of the collective excitations of arbitrary wave number due to their inelastic coupling to the fermionic quasiparticles in the collisionless regime. This rate is the main contribution to the lifetime of the collective excitations at the usual experimental temperatures \cite{bosonsfermions} when the concave collective branch \cite{Concavite} forbids the Beliaev-Landau damping mechanism \cite{Annalen}. It can be measured by Bragg spectroscopy, a technique recently applied to Fermi gases \cite{Vale2017}. With the microscopic expression of the coupling between three bosonic excitations obtained by Ref.~\cite{Annalen} and the present result, all three-body couplings between quasiparticles are known microscopically, which paves the way to a study of the dressing of quasiparticles by the interactions among them.


\section{Cold Fermi gas}


We consider a gas of neutral atomic fermions of mass $m$ interacting {via} the van der Waals force. The atoms are equally distributed in two internal states labelled by $\uparrow$ and $\downarrow$ and evolve in a homogeneous cubic space of volume $L^3$. If the gas is cold and dilute enough, the atoms interact significantly only two by two, in the $s$-wave channel, which means that the interactions among same spin fermions are negligible. In this regime, the exact form of the potential does not influence the macroscopic physics, allowing us to select an effective contact potential
\be
V(\rr,\rr')=\frac{g_0}{l^3} \delta_{\rr,\rr'}
\ee
where $\delta$ is the Kronecker symbol, and where, rather than introducing a cutoff in Fourier space, we choose to discretise space into a cubic lattice of step $l$. The bare coupling constant $g_0$ is related to the $s$-wave scattering length $a$, the experimentally accessible parameter, through the renormalisation relation \cite{Varenna}
\be
\frac{1}{g_0}=\frac {m}{4 \pi \hbar^2 a}-\int_{[-\pi/l,\pi/l[^3} \frac{\dd^3 k}{(2\pi)^3} \frac{m}{{\hbar^2 k^2}}.
\label{eq:g0_a}
\ee
In second quantisation, the grand canonical Hamiltonian of the system reads
\begin{multline}
\hat{H}= l^3 \sum_{\rr,\sigma=\uparrow/\downarrow} \hat{\psi}_\sigma^\dagger(\rr) \left( \frac{\PP^2}{2m}   -\mu\right)  \hat{\psi}_\sigma(\rr) \\
 + g_0 l^3 \sum_{{\rr}} \hat{\psi}_\uparrow^\dagger({\rr}) \hat{\psi}_\downarrow^\dagger(\rr) \hat{\psi}_\downarrow(\rr){\hat{\psi}_\uparrow}(\rr),
 \label{eq:hamiltonien_reel}
 \end{multline}
where $\PP$ is the momentum operator, whose eigenfunctions $\eee^{\ii\kk\cdot\rr}$ have eigenvalues $\hbar\kk$. 


\section{BCS theory and fermionic quasiparticles}

In this section, we recall how the fermionic excitation branch is described in the Bardeen-Cooper-Schrieffer (BCS) theory \cite{BCS1957}. We introduce the BCS Hamiltonian
\be
\hat{H}_{\rm BCS}= l^3 \sum_{\rr} \bbcro{ \sum_{\sigma=\uparrow,\downarrow}\hat{\psi}_{\sigma}^\dagger \bb{  \frac{\PP^2}{2m}   -\mu}\hat{\psi}_{\sigma}  \vphantom{\sum_\sigma} + \bb{\Delta  \hat{\psi}_\uparrow^\dagger \hat{\psi}_\downarrow^\dagger + \textrm{h.c}}}
\label{eq:HBCS}
\ee
where in the interaction term we have replaced the quadratic quantum field $g_0 \hat{\psi}_\downarrow \hat{\psi}_\uparrow$ by a self-consistent order parameter $\Delta$. The resulting quadratic Hamiltonian is readily diagonalised in Fourier space by the Bogoliubov rotation:
\be
\begin{pmatrix} \psi_\uparrow \\ \psi_\downarrow^\dagger  \end{pmatrix} = \frac{1}{L^{3/2}}\sum_{\kk\in\mathcal{D}} \eee^{\ii\kk\cdot\rr} \bbcro{ \hat{\gamma}_{\kk,\uparrow} \begin{pmatrix} U_k \\ V_k  \end{pmatrix} + \hat{\gamma}_{-\kk, \downarrow}^\dagger \begin{pmatrix} -V_k \\ U_k  \end{pmatrix}}
\label{eq:devfermions}
\ee
where $\mathcal{D}=\frac{2\pi}{L}\mathbb{Z}^{3}\cap[-\pi/{l},\pi/{l}[^3$ is the set of wave vectors of the first Brillouin zone compatible with the periodic boundary conditions.
This change of basis introduces fermionic operators $\hat{\gamma}_{\kk\sigma}^\dagger$ ($\sigma=\uparrow/\downarrow$) that create excitations whose energy $\epsilon_k$ depends on $\Delta$ and on the dispersion relation of the normal gas $\xi_k= {\hbar^2k^2}/{2m}-\mu$:
\be
\epsilon_k=\sqrt{\Delta^2+\xi_k^2}
\label{eq:epsilonk}
\ee
The $U_k$ and $V_k$ coefficients of the Bogoliubov rotation are given by
\be
U_k=\sqrt{\frac{1}{2}\bb{1+\frac{\xi_k}{\epsilon_k}}} \qquad V_k=\sqrt{\frac{1}{2}\bb{1-\frac{\xi_k}{\epsilon_k}}} 
\ee
\resoumis{and the self-consistent relation giving the value of $\Delta$ is
\be
\Delta=-\frac{g_0}{L^3}\sum_{\kk\in\mathcal{D}}\frac{\Delta}{2\epsilon_k}
\label{eq:Delta}
\ee}
The ground state of the BCS Hamiltonian \eqref{eq:HBCS} is a state in which all fermions are paired
\be
\ket{\psi_0} = \prod_{\kk\in\mathcal{D}} \bb{U_k-V_k \hat{a}_{\kk,\uparrow}^\dagger \hat{a}_{-\kk,\downarrow}^\dagger} \ket{0}
\label{eq:fondaBCS}
\ee
with a probability $|V_k|^2$ of finding the $\kk\uparrow/-\kk\downarrow$ pair. We have introduced the Fourier transform of the fermionic field operator $\hat{a}_{\kk\sigma}=\frac{1}{L^{3/2}}\sum_\kk \psi_{\sigma}(\rr)\eee^{-\ii\kk\cdot\rr}$.
The action of the operator $\hat{\gamma}_{\kk\sigma}^\dagger$ on the BCS ground state \eqref{eq:fondaBCS} is to destroy the $\kk\sigma/-\kk\sigma'$ pair and replace it by an unpaired $\kk\sigma$ fermion. This is why the fermionic excitations are referred to as \textit{pair-breaking excitations}.
In terms of those $\hat{\gamma}$ excitations, the BCS Hamiltonian is diagonal
\be
\hat{H}_{\rm BCS}=E_0+\sum_{\kk\in\mathcal{D}}\epsilon_k \bb{ \hat{\gamma}_{\kk\uparrow}^\dagger \hat{\gamma}_{\kk\uparrow} + \hat{\gamma}_{\kk\downarrow}^\dagger \hat{\gamma}_{\kk\downarrow} }
\label{eq:HBCSdiago}
\ee

\section{Description of the bosonic excitations in the RPA}


The idea of the RPA is to linearise, in the Heisenberg picture, the equations of motion of the quadratic fermionic field operators $\hat\psi_\sigma^\dagger \hat\psi_\sigma$, $\hat\psi_\uparrow^\dagger \hat\psi_\downarrow^\dagger$ and $ \hat\psi_\downarrow \hat\psi_\uparrow$ by performing incomplete Wick contractions on the quartic terms:
\begin{multline}
\hat{a} \hat{b} \hat{c} \hat{d} \rightarrow \hat{a}\hat{b} \langle\hat{c}\hat{d} \rangle  + \langle\hat{a}\hat{b} \rangle\hat{c}\hat{d} -\hat{a}\hat{c} \langle\hat{b}\hat{d}\rangle -  \langle\hat{a}\hat{c}\rangle \hat{b}\hat{d} + \hat{a}\hat{d} \langle\hat{b}\hat{c}\rangle \\ +  \langle\hat{a}\hat{d}\rangle\hat{b}\hat{c}   -  \langle\hat{a}\hat{b} \rangle  \langle\hat{c}\hat{d} \rangle + \langle\hat{a}\hat{c} \rangle  \langle\hat{b}\hat{d} \rangle- \langle\hat{a}\hat{d} \rangle  \langle\hat{b}\hat{c} \rangle
\end{multline}
where the average value $\meanv{\ldots}$ is taken in the BCS ground state \eqref{eq:fondaBCS}. Contrarily to Anderson \cite{Anderson1958}, we choose to express the resulting linear system in the basis of the fermionic quasiparticle operators $\hat{\gamma}$, by introducing the variables
\bea
\hat{s}_\kk^\qq &=& \hat{\gamma}_{-(\kk+\frac{\qq}{2}),\downarrow} \hat{\gamma}_{\kk-\frac{\qq}{2},\uparrow}    +    \hat{\gamma}_{\kk+\frac{\qq}{2},\uparrow}^\dagger \hat{\gamma}_{-(\kk-\frac{\qq}{2}),\downarrow}^\dagger \\
\hat{y}_\kk^\qq &=& \hat{\gamma}_{-(\kk+\frac{\qq}{2}),\downarrow} \hat{\gamma}_{\kk-\frac{\qq}{2},\uparrow}    -     \hat{\gamma}_{\kk+\frac{\qq}{2},\uparrow}^\dagger \hat{\gamma}_{-(\kk-\frac{\qq}{2}),\downarrow}^\dagger
\eea
In these notations, the $\qq$ coordinate is the centre-of-mass wave vector of the pair of quasiparticles and $\kk$ is the wave vector of its relative motion. In this basis, the RPA equations take the remarkably simple form \footnote{Here we neglected the Wick contractions in the Hartree-Fock channel $\hat{a}_{\kk_1\uparrow}^\dagger \hat{a}_{\kk_2\downarrow}^\dagger \hat{a}_{\kk_2'\downarrow} \hat{a}_{\kk_1'\uparrow}\to \meanv{\hat{a}_{\kk_1\uparrow}^\dagger \hat{a}_{\kk_1'\uparrow}} \hat{a}_{\kk_2\downarrow}^\dagger \hat{a}_{\kk_2'\downarrow} + \meanv{\hat{a}_{\kk_2\downarrow}^\dagger \hat{a}_{\kk_2'\downarrow}} {\hat{a}_{\kk_1\uparrow}^\dagger \hat{a}_{\kk_1'\uparrow}} $ whose contributions to the spectrum of the bosonic excitations \cite{TheseHK} and to the coupling between quasiparticles \cite{Annalen} vanish \resoumis{when the continuus space limit $l\to0$ is taken at fixed scattering length $a$}.}
\be
\!\!\!\ii\hbar \frac{\dd \hat{s}_\kk^\qq}{\dd t}  \!=\! [\epsilon_{\kk+\frac{\qq}{2}}+\epsilon_{\kk-\frac{\qq}{2}}] \hat{y}_\kk^\qq + \frac{g_0}{L^3} \! \sum_{\kk'\in\mathcal{D}} W_{\kk\qq}^+ W_{\kk'\qq}^+ \hat{y}_{\kk'}^\qq + \hat{S}_\kk^\qq 
\label{eq:RPA1}
\ee
\vspace{-0.5cm}
\be
\!\!\!\ii\hbar \frac{\dd \hat{y}_\kk^\qq}{\dd t}  \!=\! [\epsilon_{\kk+\frac{\qq}{2}}+\epsilon_{\kk-\frac{\qq}{2}}] \hat{s}_\kk^\qq +  \frac{g_0}{L^3} \! \sum_{\kk'\in\mathcal{D}} W_{\kk\qq}^- W_{\kk'\qq}^- \hat{s}_{\kk'}^\qq + \hat{Y}_\kk^\qq
\label{eq:RPA2}
\ee
\vspace{-0.7cm}
\be
\!\!\!\ii\hbar \frac{\dd}{\dd t} \hat{\gamma}_{\kk+\frac{\qq}{2},\sigma}^\dagger \hat{\gamma}_{\kk-\frac{\qq}{2},\sigma} = (\epsilon_{\kk-\frac{\qq}{2}}-\epsilon_{\kk+\frac{\qq}{2}}) \hat{\gamma}_{\kk+\frac{\qq}{2},\sigma}^\dagger \hat{\gamma}_{\kk-\frac{\qq}{2},\sigma} 
\label{eq:RPA3}
\ee
where we have introduced the cleverly chosen combinations of the Bogoliubov coefficients $U$ and $V$ already introduced in Refs.~\cite{Randeria1997,Annalen}
\be
W_{\kk\qq}^\pm=U_{\kk+\frac{\qq}{2}}U_{\kk-\frac{\qq}{2}} \pm V_{\kk+\frac{\qq}{2}}V_{\kk-\frac{\qq}{2}}
\label{eq:W}
\ee
Before giving the expressions of the source terms $\hat{S}_\kk^\qq$ and $\hat{Y}_\kk^\qq$, let us briefly comment the linear system (\ref{eq:RPA1}--\ref{eq:RPA3}). The choice of writing the equations in Fourier space has decoupled operators with a different center-of-mass wave vector $\qq$. Then the choice of singling out the quasiparticle basis has decoupled the evolution \eqref{eq:RPA3} of the quasiparticle-hole operators $\hat{\gamma}^\dagger \hat{\gamma}$, and reduced it to the trivial one given by the BCS Hamiltonian \eqref{eq:HBCSdiago}. These operators still enter into the equations of motion of $\hat{s}_\kk^\qq$ and $\hat{y}_\kk^\qq$ as source terms
\be
\hat{S}_\kk^\qq \!=\! - \frac{g_0}{L^3} W_{\kk\qq}^+ \!\!\sum_{\kkPin} \!\!  w_{\kk'\qq}^- \sum_{\sigma=\uparrow,\downarrow} {\hat{\gamma}_{\kk'+\frac{\qq}{2},\sigma}^\dagger \hat{\gamma}_{\kk'-\frac{\qq}{2},\sigma}}
\label{eq:source1}
\ee
\be
\hat{Y}_\kk^\qq =\frac{g_0}{L^3} W_{\kk\qq}^- \!\!\sum_{\kkPin} \!\!  w_{\kk'\qq}^+ \sum_{\sigma=\uparrow,\downarrow}{\hat{\gamma}_{\kk'+\frac{\qq}{2},\sigma}^\dagger \hat{\gamma}_{\kk'-\frac{\qq}{2},\sigma}}
\label{eq:source2}
\ee
where we complete the notations introduced in \eqref{eq:W}
\be
w_{\kk\qq}^\pm=U_{\kk+\frac{\qq}{2}}V_{\kk-\frac{\qq}{2}} \pm V_{\kk+\frac{\qq}{2}}U_{\kk-\frac{\qq}{2}}
\ee
We temporarily put these source terms aside to focus on the homogeneous system in $\hat{s}_\kk^\qq$ and $\hat{y}_\kk^\qq$. This system coincides exactly with the one obtained by Ref.~\cite{Annalen} in a semi-classical approach \cite{Ripka1985}. We briefly recall the results obtained by this reference on the collective excitation branch. The eigenenergy of the collective mode is determined by the implicit equation \cite{CKS2006}
\be
I_{++}(\omega_\qq,\qq) I_{--}(\omega_\qq,\qq) = {\hbar^2\omega^2_\qq} \left[I_{+-}(\omega_\qq,\qq)\right]^2
\label{eq:dedispersion}
\ee
where the $I_{\sigma\sigma'}$ are sums over the internal degrees of freedom of the pairs
\bea
\!\!\!\! I_{\pm\pm}(\omega,q) \!\!&=&\!\! \sum_{\kkin}\left[\frac{(\epsilon_{\kk+\frac{\qq}{2}}+\epsilon_{\kk-\frac{\qq}{2}})(W_{\kk\qq}^\pm)^2}{(\hbar\omega)^2
-(\epsilon_{\kk+\frac{\qq}{2}}+\epsilon_{\kk-\frac{\qq}{2}})^2}+\frac{1}{2\epsilon_\kk}\right]  \label{eq:Ipp} \\
\!\!\!\! I_{+-}(\omega,q) \!\!&=&\!\! \sum_{\kkin} \!\!\frac{W_{\kk\qq}^+ W_{\kk\qq}^-}{(\hbar\omega)^2-(\epsilon_{\kk+\frac{\qq}{2}}+\epsilon_{\kk-\frac{\qq}{2}})^2}
\label{eq:Ipm}
\eea
At low $q$, this energy is phononic
\be
\hbar\omega_\qq=\hbar c q + O(q^3)
\ee
with a speed of sound $c$ given by the hydrodynamic formula $mc^2=\rho\dd\mu/\dd\rho$.
The bosonic operators associated to the collective eigenmodes are expressed in terms of the fermionic quasiparticle-pair operators
\be
\hat{b}_\qq \!\! = - \!\! \sum_{\kk\in\mathcal{D}} \bbcro{M_{\kk}^{\qq} \hat{\gamma}_{-\kk+\frac{\qq}{2}\downarrow} \hat{\gamma}_{\kk+\frac{\qq}{2} \uparrow} -  N_{\kk}^{\qq} {\hat{\gamma}_{\kk-\frac{\qq}{2} \uparrow}^\dagger \hat{\gamma}_{-\kk-\frac{\qq}{2}\downarrow}^\dagger } } \label{eq:BqQuantique}
\ee
with the coefficients
\bea
M_{\kk}^{\qq}  &=&  \frac{\Delta  \bbcro{W_{\kk\qq}^+ + W_{\kk\qq}^- \sqrt{\frac{I_{++}(\omega_\qq,\qq)}{I_{--}(\omega_\qq,\qq)}}}}{{\norm_\qq}^{1/2}(\epsilon_{\kk+\frac{\qq}{2}}+\epsilon_{\kk-\frac{\qq}{2}}-\hbar\omega_\qq)} \label{eq:Mkq} \\
N_{\kk}^{\qq}  &=& - \frac{\Delta  \bbcro{W_{\kk\qq}^+ - W_{\kk\qq}^- \sqrt{\frac{I_{++}(\omega_\qq,\qq)}{I_{--}(\omega_\qq,\qq)}}}}{{\norm_\qq}^{1/2}(\epsilon_{\kk+\frac{\qq}{2}}+\epsilon_{\kk-\frac{\qq}{2}}+\hbar\omega_\qq)}  \label{eq:Nkq} 
\eea
\resoumis{equivalent to those (44,45) of Ref.\cite{Annalen} given that $I_{+-}(\omega_\qq,\qq) = - \sqrt{I_{++}(\omega_\qq,\qq) I_{--}(\omega_\qq,\qq)/\hbar\omega_\qq }$. These coefficients} are normalized by the constraint
\be
\sum_{\kkin} \bbcro{(M_\kk^\qq)^2-(N_\kk^\qq)^2} = 1
\label{eq:norma}
\ee
that sets the value of $\norm_\qq$.
Note that in the RPA we directly obtain quantum $\hat{b}_\qq$ operators, without need of the quantisation procedure described in Refs.~\cite{Ripka1985} and \cite{Annalen}.


\section{Coupling between bosonic and fermionic quasiparticles in the RPA}

\begin{figure}[htb]
\begin{center}
\includegraphics[width=0.25\textwidth]{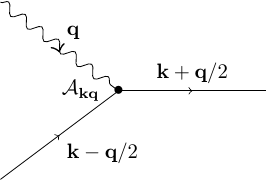}
\end{center}
\caption{\label{fig:processus} The process of absorption of a bosonic excitation (wavy line) by a BCS quasiparticle (straight line) described by the RPA.}
\end{figure}

With the work we have done in the previous section on the RPA equations, it is rather straighforward to derive the coupling between bosonic and fermionic quasiparticles. It is in fact contained in the source terms \eqref{eq:source1} and \eqref{eq:source2}. To see this, we contract the RPA system of equations \eqref{eq:RPA1} and \eqref{eq:RPA2} to form the equation of motion of the bosonic operators \eqref{eq:BqQuantique}
\be
\ii\hbar\frac{\dd \hat{b}_\qq}{\dd t}  = \hbar \omega_\qq  \hat{b}_\qq +\frac{1}{\sqrt{L^3}} \sum_{\kkin,\sigma=\uparrow,\downarrow} \mathcal{A}_{\kk\qq} \hat{\gamma}_{\kk-\frac{\qq}{2}\sigma}^\dagger \hat{\gamma}_{\kk+\frac{\qq}{2}\sigma}
\label{eq:MouvementBq}
\ee
The first term in the right hand side of this equation represents the free evolution of the bosonic quasiparticle annihilation operator, at an angular frequency $\omega_\qq$. The second describes its coupling to the fermionic operators $\hat{\gamma}$, with a coupling amplitude \footnote{\resoumis{To establish Eq.\eqref{eq:Akq} we have used the relations $\sum_\kk W^-_{\kk\qq} (M_{\kk\qq}+N_{\kk\qq})=-\frac{2L^3}{g_0}\sqrt{\frac{I_{++}(\omega_\qq,\qq)}{\mathcal{N}_\qq I_{--}(\omega_\qq,\qq) }}$ et $\sum_\kk W^+_{\kk\qq} (M_{\kk\qq}-N_{\kk\qq})=-\frac{2L^3}{g_0\sqrt{\mathcal{N}_\qq}}$ that are demonstrated from the definitions \eqref{eq:Mkq} and \eqref{eq:Nkq} of $M_{\kk\qq}$ and $N_{\kk\qq}$ using the self-consistent relation on the order parameter \eqref{eq:Delta} to recognize the sums \eqref{eq:Ipp} and \eqref{eq:Ipm}.}}
\be
\mathcal{A}_{\kk\qq} = \Delta \frac{w_{\kk\qq}^-+w_{\kk\qq}^+ \sqrt{\frac{I_{++}(\omega_\qq,q)}{I_{--}(\omega_\qq,q)}}}{\sqrt{\mathcal{N}_\qq/L^3}}
\label{eq:Akq}
\ee
that we choose independent of the system size $L^3$. We interpret the equation of motion \eqref{eq:MouvementBq} as an Heisenberg equation derived from the fictitious Hamiltonian
\begin{multline}
\hat{H}_{\rm RPA} = \sum_{\qq} \hbar \omega_{\qq} \hat{b}_\qq^\dagger \hat{b}_\qq +  \sum_{\kk,\sigma} \epsilon_\kk \hat{\gamma}_{\kk\sigma}^\dagger \hat{\gamma}_{\kk\sigma} \\ + \frac{1}{L^{3/2}} \sum_{\kk,\sigma=\uparrow,\downarrow} \mathcal{A}_{\kk\qq} \bb{ \hat{b}_\qq^\dagger \hat{\gamma}_{\kk-\frac{\qq}{2}\sigma}^\dagger \hat{\gamma}_{\kk+\frac{\qq}{2}\sigma} + \textrm{h.c.}}
\label{eq:HRPA}
\end{multline}
\resoumis{where the operators $\hat{b}_\qq$ are bosonic and commute with the fermionic operators $\hat{\gamma}_{\kk\sigma}$. The amplitude} $\mathcal{A}_{\kk\qq}$ now clearly appears as the coupling amplitude of the non-linear process represented in Fig.~\ref{fig:processus},
where a bosonic excitation is absorbed or emitted by a fermionic quasiparticle. In Sec.~\ref{sec:amortissement}, we use our knowledge of $\mathcal{A}_{\kk\qq}$ to compute the limitation of the bosonic quasiparticle lifetime caused by the process of Fig.~\ref{fig:processus}. Before that, we explain in the following section how our result \eqref{eq:Akq} can be obtained in a totally different approach based on functional integration.


\section{Comparison to the functional integral approach}

In this section we explain how the coupling between bosonic and fermionic quasiparticles can be obtained in the functional integral formalism \cite{ZinnJustin2003,Kleinert2009,Dickerscheid2009} applied to Fermi gases \cite{Randeria1997}.
The starting point is to introduce the action
\begin{multline}
S=l^3 \sum_\rr \int_0^{1/k_{\rm B}T} \dd \tau \bbacol{ \sum_{\sigma=\uparrow/\downarrow} \bar{\psi}_\sigma(\rr,\tau) \partial_\tau {\psi}_\sigma(\rr,\tau) } \\ \bbacor{\vphantom{\sum_{\sigma=\uparrow/\downarrow}}+ H[\bar{\psi}_\uparrow, {\psi}_\uparrow,\bar{\psi}_\downarrow,{\psi}_\downarrow]}
\label{eq:action}
\end{multline}
where the imaginary time $\tau=\ii t$ varies from $0$ to $1/k_{\rm B}T$, the Grassmann fields ${\psi}_\sigma$ and $\bar{\psi}_\sigma$ are the analogues of the quantum field operators $\hat{\psi}_\sigma$ and $\hat{\psi}_\sigma^\dagger$ respectively and $H[\bar{\psi}_\uparrow, {\psi}_\uparrow,\bar{\psi}_\downarrow,{\psi}_\downarrow]$ is the Hamiltonian \eqref{eq:hamiltonien_reel} where the quantum fields have been replaced by their Grassmann equivalent.
The partition function is expressed as a functional integral of this action
\be
\mathcal{Z}= \int \mathcal{D}\psi \textrm{exp} \bbcro{-S}
\label{eq:Z}
\ee
where the integral $ \int \mathcal{D}\psi$ spans all the possible configurations of the Grassmann fields. This functional integral of a quartic action cannot be performed analytically, forcing us to look for an approximation. Two disctinct yet equivalent approaches have been proposed:
(i) In the gaussian fluctuation approach, one introduces an auxiliary complex field $\Delta(\rr,\tau)$, integrates out the Grassmann fields with the Hubbard-Stratonovitch  transformation \cite{Randeria1997,Randeria2008}, then expands $\Delta(\rr,\tau)$ around the saddle point of the integral \eqref{eq:Z} so as to include the gaussian fluctuations of the action in the calculation of $\mathcal{Z}$. (ii) In the Nozières Schmidt-Rink approach, one expands the quartic part of the fermionic action \eqref{eq:action} around a zeroth order action corresponding to the BCS Hamiltonian \eqref{eq:HBCS} then resums \cite{Nozieres1985,Drummond2006} an infinity of well chosen diagrams (sometimes called ladder diagrams).
So far these approximations have been used to compute corrections to the zero temperature BCS equation of state \cite{Drummond2006,Randeria2008}, and the spectrum of the bosonic excitations \cite{Randeria2008,Tempere2011,Salasnich2015}. In this article, we show that they also describe the coupling between fermionic and bosonic excitations.

Our starting point is the equation giving the eigenenergy of the bosonic modes obtained by Ref.~\cite{Randeria1997}
\be
\textrm{det}\, M(z_\qq,\qq)=0
\label{eq:detM}
\ee
where $z_\qq$ is the eigenenergy and $M$, sometimes called the gaussian fluctuation matrix, is a $2\times2$ matrix whose coefficients, after correction \cite{Randeria2008} of the sign mistake in Ref.~\cite{Randeria1997}, are given by $M_{22}(z,\qq)=M_{11}(-z,-\qq)$, $M_{12}=M_{21}$ and
\begin{widetext}
\bea
\!\!\!\!\!\!M_{11}(z,\qq) \!\!\!&=&\!\!\! -\frac{L^3}{g_0} + \sum_{\kk} (1-n_+^{\rm F}-n_-^{\rm F}) \bbcro{\frac{U_+^2 U_-^2}{z-\epsilon_+ -\epsilon_-}-\frac{V_+^2 V_-^2}{z+\epsilon_+ +\epsilon_-}}
\!\!+\!\!\sum_{\kk} (n_+^{\rm F}-n_-^{\rm F}) \bbcro{\frac{V_+^2 U_-^2}{z+\epsilon_+ -\epsilon_-}-\frac{U_+^2 V_-^2}{z-\epsilon_+ +\epsilon_-}} \label{eq:M11} \\
\!\!\!\!\!\!M_{12}(z,\qq) \!\!\!&=&\!\!\! \sum_{\kk} (1-n_+^{\rm F}-n_-^{\rm F}) \bbcro{\frac{U_+ U_- V_+ V_- }{z+\epsilon_+ + \epsilon_-} - \frac{U_+ U_- V_+ V_- }{z-\epsilon_+ -\epsilon_-}}
+\sum_{\kk} (n_+^{\rm F}-n_-^{\rm F}) \bbcro{\frac{U_+ V_+ U_- V_- }{z+\epsilon_+ -\epsilon_-}-\frac{U_+ V_+ U_- V_- }{z-\epsilon_+ + \epsilon_-}} \label{eq:M12}
\eea
\end{widetext}
where we permit ourselves the short-hand notation $a_\pm=a_{\kk\pm\frac{\qq}{2}}$ and we introduce the Fermi-Dirac occupation numbers of the fermionic quasiparticles
\be
n_\kk^{\rm F}=\frac{1}{\eee^{\epsilon_\kk/k_{\rm B}T}+1}
\ee
Notice that in (\ref{eq:M11})--(\ref{eq:M12}) the variable $z$ can take in principle only the pure imaginary values corresponding to the Matsubara frequencies $z=\ii\hbar 2 \pi n k_{\rm B}T$, $n\in\mathbb{N}$.
To relate the functional integral approach to the RPA, we do a small change of variables
\begin{multline}
M_{\pm\pm} = \frac{M_{11}+M_{22}}{2}\mp M_{12} \\ = \sum_\kk { (W_{\kk\qq}^\pm)^2 \frac{(1-n^{\rm F}_{\kk+\frac{\qq}{2}}-n^{\rm F}_{\kk-\frac{\qq}{2}})(\epsilon_{\kk+\frac{\qq}{2}} + \epsilon_{\kk-\frac{\qq}{2}})}{z^2-(\epsilon_{\kk+\frac{\qq}{2}} + \epsilon_{\kk-\frac{\qq}{2}})^2}} \\ - \sum_\kk {(w_{\kk\qq}^\mp)^2 \frac{(n^{\rm F}_{\kk+\frac{\qq}{2}} - n^{\rm F}_{\kk-\frac{\qq}{2}})(\epsilon_{\kk+\frac{\qq}{2}} - \epsilon_{\kk-\frac{\qq}{2}})}{z^2-(\epsilon_{\kk+\frac{\qq}{2}} - \epsilon_{\kk-\frac{\qq}{2}})^2}} -\frac{L^3}{g_0}  \label{eq:Mpp}
\end{multline}
\begin{multline}
M_{+-}  = \frac{M_{11}-M_{22}}{2z} \\ =   \sum_\kk { W_{\kk\qq}^+ W_{\kk\qq}^- \frac{1-n^{\rm F}_{\kk+\frac{\qq}{2}}-n^{\rm F}_{\kk-\frac{\qq}{2}}}{z^2-(\epsilon_{\kk+\frac{\qq}{2}} + \epsilon_{\kk-\frac{\qq}{2}})^2}} \\
- \sum_\kk{w_{\kk\qq}^+w_{\kk\qq}^- \frac{n^{\rm F}_{\kk+\frac{\qq}{2}}-n^{\rm F}_{\kk-\frac{\qq}{2}}}{z^2-(\epsilon_{\kk+\frac{\qq}{2}} - \epsilon_{\kk-\frac{\qq}{2}})^2}} 
\label{eq:Mpm}
\end{multline}
These new sums coincide at zero temperature with those we introduced in the RPA \footnote{At zero temperature we analytically continue the $M_{\sigma\sigma'}$ functions to the real axis by setting $z\mapsto\hbar\omega$, which we can do without trouble as long as $\hbar\omega$ stays below the gapped $\kk\mapsto\epsilon_{\kk+\frac{\qq}{2}}+\epsilon_{\kk-\frac{\qq}{2}}$ continuum.}
\be
M_{\sigma\sigma'} \underset{T\to0}{\to} I_{\sigma\sigma'}, \quad \sigma,\sigma'=\pm
\ee
Moreover, they fulfill a similar eigenvalue equation
\be
M_{++}(z_\qq,\qq) M_{--}(z_\qq,\qq) = {z_\qq^2} \left[M_{+-}(z_\qq,\qq)\right]^2
\label{eq:dedispersionTneq0}
\ee
We chose the variables \eqref{eq:Mpp} and \eqref{eq:Mpm} also for a practical reason, as they substantially simplify the long wavelength calculations (see Sec.~\ref{sec:petitsq}).

\resoumis{Since Eq.\eqref{eq:dedispersionTneq0} has no solution on the imaginary axis, it is natural to extend the definition of the $M_{\sigma\sigma'}$  functions to the complex plane by setting
\be
z=\hbar\omega- \frac{\ii\hbar\Gamma}{2}
\ee
where the factor $2$ is chosen so that $\Gamma$ corresponds to the damping rate of the mode. This is not a problem at zero temperature (that is when $n_\kk^{\rm F}=0$, $\forall \kk\in\mathcal{D}$) and this leads to the eigenvalue equation \eqref{eq:dedispersion} already encountered in the framework of the RPA, which for fixed $\qq$ possesses a unique real solution $\hbar\omega_\qq<\textrm{min}_{\kk}(\epsilon_{\kk+\frac{\qq}{2}}+\epsilon_{\kk-\frac{\qq}{2}})$. At non zero temperature, when $n_\kk^{\rm F}\neq0$, $\forall \kk\in\mathcal{D}$, the situation is more complicated: the function $z\mapsto M_{++}(z,\qq) M_{--}(z,\qq) -{z^2} \left[M_{+-}(z,\qq)\right]^2$ has a branch cut along the whole real axis, corresponding to the continuum $\kk\mapsto\epsilon_{\kk+\frac{\qq}{2}}-\epsilon_{\kk-\frac{\qq}{2}}$ (which in the RPA we view as the eigenenergy of the quasiparticle-hole operator $\hat{\gamma}_{\kk-\frac{\qq}{2}}^\dagger\hat{\gamma}_{\kk+\frac{\qq}{2}}$ and which clearly reaches zero $0$ when $\qq=0$) appearing in the denominators of the last lines of \eqref{eq:Mpp} and \eqref{eq:Mpm}. This function has then a discontinuity when crossing the real axis and no root outside of it : Eq.\eqref{eq:dedispersionTneq0}, as such, has no solution.}

\resoumis{However, as in the case of a discret state coupled to a continuum \cite{CohenChapIIIC}, the inverse function $z\mapsto (M_{++}(z,\qq) M_{--}(z,\qq) -{z^2} \left[M_{+-}(z,\qq)\right]^2)^{-1}$ retains the memory of its pole at zero temperature in the shape of its variations in the vicinity of the branch cut. In the perturbative regime, when the coupling to the continuum is weak, these variations are fast around $z=\hbar\omega_\qq$ and denote the existence of a pole $z_\qq$ in the analytic continuation of the inverse function across the branch cut, for $\textrm{Im}z<0$. Here we restrict to this perturbative regime and look for the new solution $z_\qq$ of the analytically continuated Eq.\eqref{eq:dedispersionTneq0} in the vicinity of the zero temperature solution $\hbar\omega_\qq$:}
\be
|z_\qq-\hbar\omega_\qq| \ll \hbar\omega_\qq
\ee
\resoumis{Clearly,} this regime is reached for sufficiently low temperatures. More precisely, we assume that the deviations to the zero temperature spectrum behave as
\be
|z_\qq-\hbar\omega_\qq|=O(\eee^{-\Delta/k_{\rm B}T})
\ee
We justify this hypothesis (which the final result will also confirm) by the fact that $M_{\sigma\sigma'}$ depends on temperature through the fermionic occupation numbers $n_\kk^{\rm F}\simeq\eee^{-\epsilon_\kk/k_{\rm B}T}$ at low temperature. 
To expand $M_{\sigma\sigma'}$ in powers of $|z_\qq-\hbar\omega_\qq|$, we split the contributions of the second and third lines of \eqref{eq:Mpp} and \eqref{eq:Mpm}. The terms of the second lines have no branch cut that reaches $0$, and are therefore straightforwardly continued to the lower half-plane : we simply expand in powers of $|z_\qq-\hbar\omega_\qq|$ inside the summation. The terms of the third lines are those whose branch cut reaches $0$ but they are already of order $O(\eee^{-\beta\Delta})$ (because of the $n^{\rm F}_{\kk+\frac{\qq}{2}}-n^{\rm F}_{\kk-\frac{\qq}{2}}$ factor): it is enough to obtain their value for $|z_\qq-\hbar\omega_\qq|\to0$, which we do by approaching the branch cut from above setting  $z=\hbar\omega_\qq+\ii\eta$ with $\eta\to0^+$.
In this article, we focus on the imaginary part $\Gamma_\qq=O(\eee^{-\beta\Delta})$ of the correction to the spectrum. To obtain it, it is enough to expand the real and imaginary parts of  $M_{\sigma\sigma'}$ up to first non-zero order in $\eee^{-\Delta/k_{\rm B}T}$:
\bea
\textrm{Re}\,M_{\sigma\sigma'}(z_\qq,\qq) \!\!\!&=&\!\!\! I_{\sigma\sigma'}(\omega_\qq,\qq) + O(\eee^{- \Delta/k_{\rm B}T}) \label{eq:ReM} \\
\textrm{Im}\,M_{\sigma\sigma'}(z_\qq,\qq) \!\!\!&=&\!\!\!  \Gamma_\qq \, K_{\sigma\sigma'}(\omega_\qq,\qq) + J_{\sigma\sigma'}(\omega_\qq,\qq) \label{eq:ImM} \\ && + O\bb{[\eee^{- \Delta/k_{\rm B}T}]^2} \notag 
\eea
where in Eq.~\eqref{eq:ImM} the $K_{\sigma\sigma'}$ and $J_{\sigma\sigma'}$ contributions stem respectively from the second and third lines of \eqref{eq:Mpp} and \eqref{eq:Mpm}. Explicitly, we have
\bea
K_{\pm\pm} (\omega,\qq) &=& \hbar\omega \sum_\kk  \frac{ (W_{\kk\qq}^\pm)^2 (\epsilon_{\kk+\frac{\qq}{2}} + \epsilon_{\kk-\frac{\qq}{2}})}{[\hbar^2\omega^2-(\epsilon_{\kk+\frac{\qq}{2}} + \epsilon_{\kk-\frac{\qq}{2}})^2]^2} \label{eq:Kpp} \\
K_{+-} (\omega,\qq) &=& \hbar\omega \sum_\kk \frac{ W_{\kk\qq}^+ W_{\kk\qq}^-}{[\hbar^2\omega^2-(\epsilon_{\kk+\frac{\qq}{2}} + \epsilon_{\kk-\frac{\qq}{2}})^2]^2} \label{eq:Kpm} \\
J_{\pm\pm} (\omega,\qq) &=& - {\pi} \sum_\kk (w^\mp_{\kk\qq})^2 {(n^{\rm F}_{\kk+\frac{\qq}{2}}-n^{\rm F}_{\kk-\frac{\qq}{2}})} \label{eq:Jpp} \\ &&\qquad\qquad\times {\delta(\hbar\omega+\epsilon_{\kk+\frac{\qq}{2}}-\epsilon_{\kk-\frac{\qq}{2}}) } \notag  \\
J_{+-} (\omega,\qq) &=&  \frac{\pi}{\hbar\omega} \sum_\kk w^+_{\kk\qq} w^-_{\kk\qq} {(n^{\rm F}_{\kk+\frac{\qq}{2}}-n^{\rm F}_{\kk-\frac{\qq}{2}})}  \label{eq:Jpm} \\ &&\qquad\qquad\times {\delta(\hbar\omega+\epsilon_{\kk+\frac{\qq}{2}}-\epsilon_{\kk-\frac{\qq}{2}})} \notag
\eea
Inserting the expansions \eqref{eq:ReM} and \eqref{eq:ImM} in the eigenvalue equation \eqref{eq:dedispersionTneq0} yields the expression of the damping rate
\be
\hbar\Gamma_\qq = 4\Delta^2 \bbcro{\frac{2\hbar^2\omega_\qq^2 J_{+-}I_{+-}-J_{++}I_{--}-J_{--}I_{++}}{I_{--} \mathcal{N}_\qq}} \label{eq:Gammaq1}\\
\ee
where all the $J_{\sigma\sigma'}$ and $I_{\sigma\sigma'}$ functions are evaluated in $(\omega_\qq,\qq)$. To identify the normalisation constant $\mathcal{N}_\qq$ of the RPA, we have used the relation
\be
\frac{\mathcal{N}_\qq}{4 \Delta^2} =  \frac{ [K_{++} I_{--}+K_{--} I_{++}-2\hbar^2\omega_\qq^2  I_{+-} K_{+-}+\hbar\omega_\qq I_{+-}^2]}{I_{--}} 
\label{eq:norma2}
\ee
which is shown in the framework of the RPA by replacing expressions \eqref{eq:Mkq} and \eqref{eq:Nkq} in the normalisation condition \eqref{eq:norma} and then using the zero temperature eigenvalue equation \eqref{eq:dedispersion}. We conclude this calculation by replacing in \eqref{eq:Gammaq1} the $J_{\sigma\sigma'}$ functions by their expressions (\ref{eq:Jpp}) and (\ref{eq:Jpm})
\be
\hbar\Gamma_\qq= 4\pi \sum_\kk {\frac{\mathcal{A}^2_{\kk\qq}}{L^3} (n^{\rm F}_{\kk-\frac{\qq}{2}}-n^{\rm F}_{\kk+\frac{\qq}{2}}) \delta(\hbar\omega_\qq+\epsilon_{\kk-\frac{\qq}{2}}-\epsilon_{\kk+\frac{\qq}{2}})} \label{eq:Gammaq2}
\ee
This expression of $\Gamma_\qq$ makes an elegant connection with the RPA expression of the coupling amplitude $\mathcal{A}_{\kk\qq}$: it is the damping rate one obtains by applying the Fermi golden rule to the population of the bosonic excitations of wave vector $\qq$ using the RPA Hamiltonian \eqref{eq:HRPA} and adding up the damping rates due to the $\uparrow$ and $\downarrow$ fermionic excitations, here identical. Expression \eqref{eq:Gammaq2} is valid only to leading order in $\eee^{-\Delta/k_{\rm B}T}$, so we should keep only the leading term in the fermionic occupation numbers
\be
n^{\rm F}_{\kk}=\eee^{-\epsilon_\kk/k_{\rm B}T}+O\bb{[\eee^{-\epsilon_\kk/k_{\rm B}T}]^2}
\ee
Our perturbative approximation therefore assumes that the gas of fermionic quasiparticles is non-degenerate.


\section{Phonon damping}
\label{sec:amortissement}
In this section, we compute an explicit expression of the phonon damping rate \eqref{eq:Gammaq2} due to the coupling to the fermionic quasiparticles. We first concentrate on the long wavelength limit, in which our result can be compared to other approaches.


\subsection{Long wavelength limit}
\label{sec:petitsq}
\begin{figure}
\begin{center}
\includegraphics[width=0.49\textwidth]{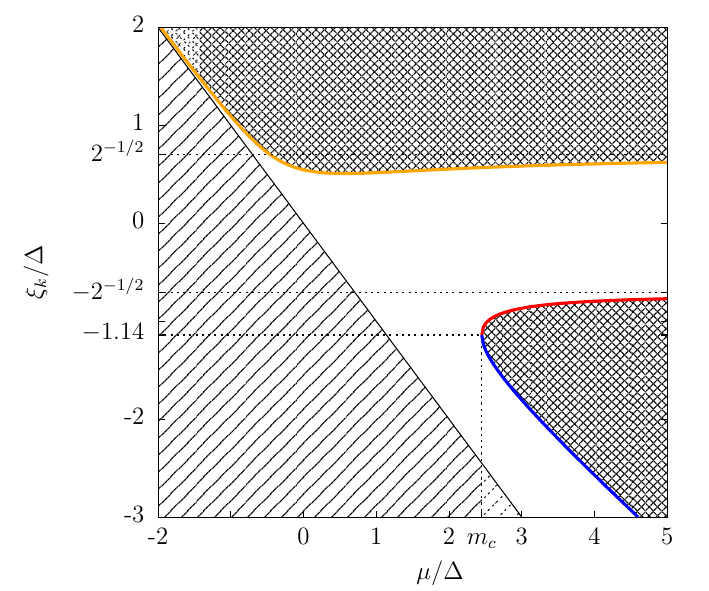}
\end{center}
\caption{\label{fig:Bornes} The integration domain $\mathcal{X}$ over the variable $\xi=\xi_k/\Delta$ is shown in the long wavelength limit as a function of $\mu/\Delta$. The BEC limit is found at $\mu/\Delta\to-\infty$ and the BCS limit at $\mu/\Delta\to+\infty$. Without any constraint, the integration domain is physically restricted to $\xi>-\mu/\Delta$, that is outside the hatched region. For all values of $\mu/\Delta$, there exists an integration domain $\mathcal{X}_1\subset\mathbb{R}^+$ (upper part of the graph) in the increasing part of the BCS branch; this interval is bounded from below by $\xi_1$ (full orange line) and not bounded from above. On the BCS side, for $\mu/\Delta>m_c\simeq2.45$ (or $1/k_{\rm F}a<-0.594$), there exists a second interval $\mathcal{X}_2\subset\mathbb{R}^-$ in the decreasing part of the BCS branch, bounded from below by $\xi_2$ (full blue line) and from above by $\xi_3$ (full red line). Note that the vicinity of the minimum of the BCS branch (at $\xi=0$) is always outside the integration domain.}
\end{figure}

In the long wavelength limit, $q\to0$ (that is both $\hbar^2q^2/m\Delta\to0$ and $\hbar \omega_\qq/k_{\rm B}T\to0$), we first simplify in Eq.~\eqref{eq:Gammaq2} the argument of the Dirac delta function expressing energy conservation
\be
\hbar\omega_\qq+\epsilon_{\kk-\frac{\qq}{2}}-\epsilon_{\kk+\frac{\qq}{2}} = \hbar\omega_\qq\bbcro{1-\frac{\xi_k}{\epsilon_k}\frac{\hbar k}{mc}u+O(q^2)} =0
\label{eq:consE}
\ee
where we denote the cosine of the angle between $\kk$ and $\qq$ by
\be
u=\frac{\kk\cdot\qq}{kq}
\ee
There exists a value of $u$ satisfying the energy conservation constraint \eqref{eq:consE} if and only if 
\be
\frac{\xi_k^2}{\epsilon_k^2}\frac{\hbar^2 k^2}{m^2c^2} >1
\ee
Physically, this constraint means that the absorption of a phonon is allowed only if the slope $\frac{\dd\epsilon_k}{\hbar\dd k}=\frac{\xi_k}{\epsilon_k} \frac{\hbar k}{m}$ of the fermionic branch is greater in absolute value than the speed of sound $c$. This condition may be interpreted as a Landau criterion if we view the fermionic quasiparticle as an impurity travelling in the superfluid at the group velocity $\frac{\dd\epsilon_k}{\hbar\dd k}$.
Eliminating the norm of $k$ in favor of $\xi=\xi_k/\Delta$ with the change of variable $\xi_k=\hbar^2k^2/2m-\mu$, this constraint takes a polynomial form
\be
 \xi^3 +\bb{\frac{\mu}{\Delta}-\frac{m{c}^2}{2\Delta}} \xi^2 - \frac{m{c}^2}{2\Delta} > 0
\label{eq:contrainteP}
\ee
Since $m{c}^2$ is a function of $\mu/\Delta$ through the BCS equation of state \cite{Concavite}, the coefficients of this polynomial depend in fact on only one parameter. In Fig.~\ref{fig:Bornes} we show as a function of $\mu/\Delta$ the set $\mathcal{X}$ of values of $\xi$ for which the energy conservation constraint can be fulfilled. On the Bose-Einstein Condensate (BEC) side $\mu/\Delta\to-\infty$ where the BCS branch $k\mapsto\epsilon_k$ is a strictly increasing function of $k$, $\mathcal{X}$ is of the form $[\xi_1,+\infty[$ with a lower bound $\xi_1>0$. When $\mu>0$, the BCS branch is a decreasing function for $k<(2m\mu/\hbar^2)^{1/2}$; when the slope of the branch in this region becomes larger than the speed of sound $c$, that is for $\mu/\Delta=2.45$, there appears a second contribution to $\mathcal{X}$, of the form $[\xi_2,\xi_3]$ with $\xi_2<\xi_3<0$.

We then expand in powers of $q$ expression \eqref{eq:Akq} of the coupling amplitude \footnote{\resoumis{Eq.\eqref{eq:dedispersion} yields the equivalent $\sqrt{I_{++}(\omega_\qq,\qq)/I_{--}(\omega_\qq,\qq)}{\sim}$ $\hbar c q |I_{+-}(0,\zero)/I_{--}(0,\zero)|$ when $q\to0$ and Eq.(14) of Ref.\cite{Concavite} yields $2|I_{+-}(0,0)/I_{--}(0,0)|=\dd\Delta/\Delta\dd\mu$.}}
\be
\mathcal{A}_{\kk\qq} = \bb{\frac{\hbar\omega_\qq}{2}\frac{\dd \mu}{\dd \rho}}^{1/2} \frac{\Delta}{\epsilon_k} \bb{\frac{\dd\Delta}{\dd\mu} + \frac{\Delta}{\epsilon_k}  \frac{\hbar k}{mc} u} + O(q^{5/2})
 \label{eq:DLA}
\ee
as well as the factor containing the occupation numbers
\be
n^{\rm F}_{\kk-\frac{\qq}{2}}-n^{\rm F}_{\kk+\frac{\qq}{2}} = \frac{\hbar\omega_\qq}{k_{\rm B}T} \eee^{-\epsilon_\kk/k_{\rm B}T} +O(q^{3})
\label{eq:nombresdocc1}
\ee
Last, we compute the rate $\Gamma_\qq$ in the thermodynamic limit $\frac{1}{L^3}\sum_\kk\to\frac{1}{(2\pi)^3}\int \dd^3 k$, in a spherical frame with $z$-axis along $\qq$, and using the Dirac delta function to integrate out the colatitude $u$ 
\be
\frac{\Gamma_\qq}{\omega_\qq} \!\!=\!\! \frac{3\pi}{2} \bbcro{\frac{c}{v_{\rm F}}}^3 \frac{\Delta}{k_{\rm B}T} \int_{\mathcal{X}}  \frac{\dd\xi}{\epsilon|\xi|} \bbcro{\frac{\dd\Delta}{\dd\mu}+\frac{1}{\xi}}^2 \eee^{-\Delta\epsilon/k_{\rm B}T}  +O(q^2)
\label{eq:Gammapttsq}
\ee
We have introduced the notation $\epsilon=\sqrt{\xi^2+1}$ and the Fermi velocity $v_{\rm F}$ related to the density by $3\pi^2\rho=(mv_{\rm F}/\hbar)^3$. 
\begin{figure}[htb]
\begin{center}
\includegraphics[width=0.49\textwidth]{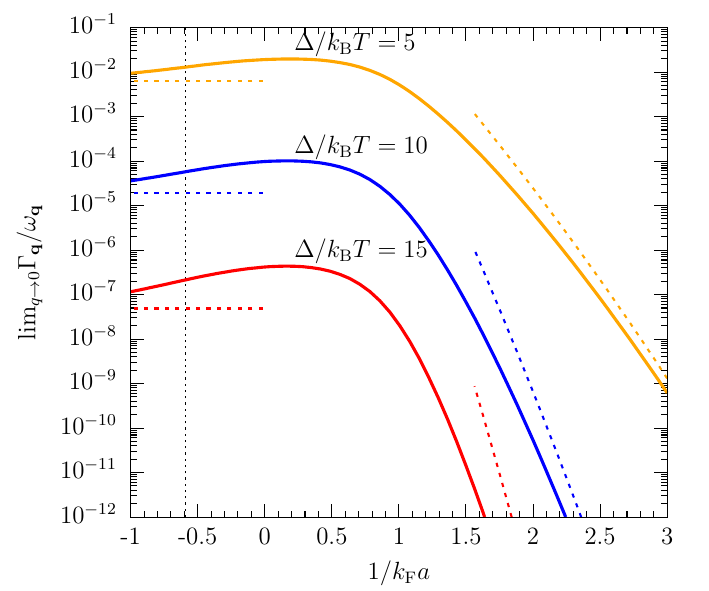}
\end{center}
\caption{Inverse quality factor $\Gamma_\qq/\omega_\qq$ of the long wavelength phonons as a function of the interaction regime (measured by $1/k_{\rm F}a$) for different values of temperature, from top to bottom, $\beta\Delta=5$, $10$ and $15$. In the BCS limit, the dashed lines show the asymptotic value for $1/k_{\rm F}a\to-\infty$ (or $\mu/\Delta\to+\infty$): $\lim_{q\to0}\Gamma_\qq/\omega_\qq\underset{1/k_{\rm F}a\to-\infty}{\to} \frac{\pi}{\sqrt{3}} f(\beta\Delta)$ with {\normalsize $f(x)=x\int_{\sqrt{3/2}}^\infty  \frac{\eee^{-xt} \dd t}{(t^2-1)^2}$}. In the BEC limit the dashed lines indicate the oblique asymptote for $1/k_{\rm F}a\to+\infty$ (or $\mu/\Delta\to-\infty$): $\lim_{q\to0}\Gamma_\qq/\omega_\qq\underset{1/k_{\rm F}a\to+\infty}{\sim} {8} (k_{\rm F}a)^{3/2} \eee^{-\mu/k_{\rm B}T}/{\sqrt{3\pi}}$. To the left of the vertical dotted line (for $1/k_{\rm F}a<-0.594$) the domain $\mathcal{X}$ contains a contribution of the decreasing part of the BCS branch.
\label{fig:Fig3}}
\end{figure} 
The ratio ${\Gamma_\qq}/{\omega_\qq}$ tends to a \resoumis{non-zero} constant at low $q$. This allows us to represent on Fig.  \ref{fig:Fig3} the long wavelength limit of ${\Gamma_\qq}/{\omega_\qq}$ as a function of the interaction strength for different values of temperature. 
As expected, this value is exponentially small in temperature because of the $\eee^{-\Delta\epsilon/k_{\rm B}T}$ factor in the integrand. At very low temperature, this exponential factor is extremely peaked around its maximum for $\xi_1=\textrm{min}\{|\xi|,\xi\in\mathcal{X}\}$ (full orange line on Fig.~\ref{fig:Bornes}), and we can replace the non exponential factor in the integrand by its value in $\xi=\xi_1$ to obtain
\be
\lim_{q\to0}\frac{\Gamma_\qq}{\omega_\qq} \underset{\substack{T\to0 \\}}{\sim} \frac{3\pi}{2} \bbcro{\frac{c}{v_{\rm F}}}^3  \bbcro{\frac{\dd\Delta}{\dd\mu}+\frac{1}{\xi_1}}^2 \frac{\eee^{-\Delta'/k_{\rm B}T} }{\xi_1^2} 
\label{eq:Gammapttsq2}
\ee
In this expression there appears an effective gap $\Delta'=\Delta\sqrt{\xi_1^2+1}$ that reflects the fact that the absorption-emission process is not resonant for excitations located at the minimum of the fermionic branch.


\subsection{Comparison to other approaches}


Several other methods exist to obtain the coupling amplitude
$\mathcal{A}_{\kk\qq}$ in the long wavelength limit. 
In their description of superfluid helium-4 with
quantum hydrodynamics, Landau and Khalatnikov \cite{Khalatnikov1949} 
derive an expression for the coupling between phonons and rotons by
treating the phonons as a semiclassical hydrodynamic perturbation acting
on a generic gapped roton Hamiltonian. Their expression
can be generalized \cite{bosonsfermions} to ultracold atomic Fermi
gases where the fermionic branch $k\mapsto\epsilon_{\textbf{k}}$
plays the role of the rotons:
\be
\mathcal{A}_{\kk\qq}^{\rm hydro} = \bb{\frac{\hbar\omega_\qq}{2}\frac{\dd \mu}{\dd \rho}}^{1/2}  \bb{\frac{\dd\epsilon_k}{\dd\mu} +  \frac{\hbar k}{mc} u } + O(q^{5/2}) 
\label{eq:Ahydro}
\ee
This expression for $\mathcal{A}$ coincides with our result
\eqref{eq:DLA} for resonant processes only (i.e. processes obeying
the energy conservation relation \eqref{eq:consE} \resoumis{which is used to eliminate the angular variable $u$}), as expected
for an effective low energy theory such as quantum hydrodynamics.
This is suitable for a calculation involving on-shell energies, such 
as that of the damping rate $\Gamma_{\textbf{q}}$. However, for
quantities depending on off-shell coupling amplitudes, such as 
energy shifts of the dispersion relation, the hydrodynamic
result \eqref{eq:Ahydro} cannot be used. 

Note that Ref.~\cite{bosonsfermions}
uses the quantum hydrodynamic theory to calculate the damping rate
of phonons due to the scattering of phonons on BCS excitations. This
is the dominant process at the lowest temperatures because 
it can resonantly couple to the excitations at the bottom of the 
fermionic branch: the associated damping rate behaves as $\eee^{-\Delta/k_{\rm B}T}$, without an effective gap as in \eqref{eq:Gammapttsq2}. However, in practice it can be much smaller than the rate $\Gamma_\qq$ at the experimentally accessible temperatures.
Obtaining four-body couplings (between two BCS excitations and two collective excitations) in a microscopic theory requires us to extend our
description beyond the RPA, or beyond the gaussian fluctuation approximation
in the functional integral framework, and this is beyond the scope of
the present paper.

Also the functional integral approach has been used previously
to calculate the damping rate $\Gamma_\qq$ in the long wavelength limit \cite{VincentLiu2011}, and
a different result was reported than what we obtain in the
present treatment, Eq.~\eqref{eq:Gammapttsq}. 
One of the reasons for this difference is that in 
Ref.~\cite{VincentLiu2011} the coefficients $K_{\sigma \sigma'}$ arising 
from the non-singular part of $M_{\sigma \sigma'}$ are neglected, based 
on the fact that the denominator
$\hbar\omega_\qq\pm(\epsilon_{\kk+\frac{\qq}{2}}+\epsilon_{\kk-\frac{\qq}{2}})$ is not resonant. Here we show
that this contribution cannot be neglected, as it is essential to
retrieve the RPA result and modifies significantly the value
of $\Gamma_{\textbf{q}}$, also at low $q$.


\subsection{Beyond the long wavelength limit}


Next, we turn to the general result \eqref{eq:Gammaq2} beyond
the long wavelength limit. As mentioned above, this limit requires 
two conditions to be met: 
$\hbar c q /k_{\rm B}T\ll1$ and $\hbar^2q^2/m\Delta\ll1$. Relaxing
the first condition is straightforward as only the
expansion of the fermionic occupations numbers \eqref{eq:nombresdocc1} uses the inequality $\hbar c q /k_{\rm B}T\ll1$.
Using energy conservation and the relations 
$(1-n^{\rm F}_\kk)/n^{\rm F}_\kk=\eee^{\beta\epsilon_\kk}$ and
$(1+n^{\rm B}_\qq)/n^{\rm B}_\qq=\eee^{\beta\hbar\omega_\qq}$ for fermionic
and bosonic occupation numbers respectively, one finds \cite{bosonsfermions}:
\be
n^{\rm F}_{\kk-\frac{\qq}{2}}-n^{\rm F}_{\kk+\frac{\qq}{2}} = \frac{\eee^{-\epsilon_{\kk-\frac{\qq}{2}}/k_{\rm B}T}}{1+{n}^{\rm B}_\qq} +O([\eee^{-\Delta/k_{\rm B}T}]^2)
\ee
with the bosonic occupation numbers
\be
n^{\rm B}_{\qq}=\frac{1}{\eee^{\hbar\omega_\qq/k_{\rm B} T}-1},
\ee
This allows us to write \eqref{eq:Gammapttsq} for arbitrary values of $\hbar c q /k_{\rm B}T$ as
\be
\hbar{\Gamma_\qq}\simeq \frac{3\pi}{2} \bbcro{\frac{c}{v_{\rm F}}}^3 \frac{\Delta}{1+n_\qq^{\rm B}} \int_{\mathcal{X}}  \frac{\dd\xi}{\epsilon|\xi|} \bbcro{\frac{\dd\Delta}{\dd\mu}+\frac{1}{\xi}}^2 \eee^{-\Delta\epsilon/k_{\rm B}T} 
\ee

Relaxing the remaining condition $\hbar^2q^2/m\Delta\ll1$ is more
difficult, as no analytic expression is available for the collective 
branch $q\mapsto\omega_\qq$ outside the long wavelength limit.
Similarly as in Refs.~\cite{CKS2006,Concavite}, we numerically
solve $\omega_{\textbf{q}}$ from the implicit equation \eqref{eq:dedispersion} and
use this result to evaluate the integral in Eq.~\eqref{eq:Gammaq2}. Details
on how to satisfy the energy conservation constraint can be found in Appendix \ref{ann:domaine}.

In figures \ref{fig:GamLU} and \ref{fig:GamNonConnexe} we plot the
inverse quality factor $\Gamma_\qq/\omega_\qq$ as a 
function of wave number $q$, for several interaction strengths, at
a temperature $\Delta/k_{\rm B}T=5$. For all interaction strengths
the inverse quality factor decreases quadratically at low $q$,
starting from its limiting value \eqref{eq:Gammapttsq} for $q\to0$. 

Figure \ref{fig:GamLU} shows the damping rate in the BCS regime 
($1/k_{\rm F}a=-0.47$) and at unitarity ($1/k_{\rm F}a=0$). For these two cases,
the collective branch $q\mapsto\omega_\qq$ disappears when hitting the continuum
of \resoumis{fermionic biexcitations} 
$\kk\mapsto\epsilon_{\kk+\frac{\qq}{2}}+\epsilon_{\kk-\frac{\qq}{2}}$
at $\hbar q/(2m\mu)^{1/2}\simeq2.0$ and $2.3$, respectively. In these
points, the inverse quality factor becomes vanishingly small along a 
vertical tangent. The suppression of $\Gamma_\qq$ is due to
a divergence in the normalization constant $\mathcal{N}_\qq$ 
\eqref{eq:norma2} which, in turn, comes from the denominators
$(\hbar\omega_\qq-\epsilon_{\kk+\frac{\qq}{2}}-\epsilon_{\kk-\frac{\qq}{2}})^2$ in
the integral quantities $K_{\sigma\sigma'}$ (\ref{eq:Kpp},\ref{eq:Kpm}), 
which becomes a second order pole as  $\omega_\qq$ enters the pair breaking
continuum. Physically, this suppression relates to the fact that when the branch
approaches the continuum, the emission and absorption processes that are considered here
 are dominated by the level repulsion exerted by the continuum.

Figure \ref{fig:GamNonConnexe} shows the interesting case, on the BEC side,
that occurs when the the collective branch $q\mapsto\omega_\qq$ is not 
singly connected \cite{CKS2006}, but consists of two pieces 
$[0,q_{\rm sup}]\cup[q_{\rm inf},+\infty[$ with $q_{\rm sup}<q_{\rm inf}$,
as shown in the inset. Now $\Gamma_{\textbf{q}}$ goes to zero in the two
points $q_{\rm sup},q_{\rm inf}$ where the collective branch reaches the continuum.
For large $q$ the collective excitations acquire an energy $\hbar^2 q^2/4 m$ 
equal to the kinetic energy of a molecule of mass $2 m$ and momentum $\hbar q$. In
Ref.~\cite{CKS2006} this is interpreted as a tightly bound molecule that is ejected out of the pair condensate. These high energy excitations can only
be absorbed by fermionic quasiparticles whose wave number $k$ is large enough
so that the maximal energy difference for aligned wavevector $\epsilon_{k+\frac{q}{2}}-\epsilon_{k-\frac{q}{2}}$
exceeds $\hbar \omega_\qq$. This explains the strong decrease of
the inverse quality factor for large $q$.

\begin{figure}
\begin{center}
\includegraphics[width=0.49\textwidth]{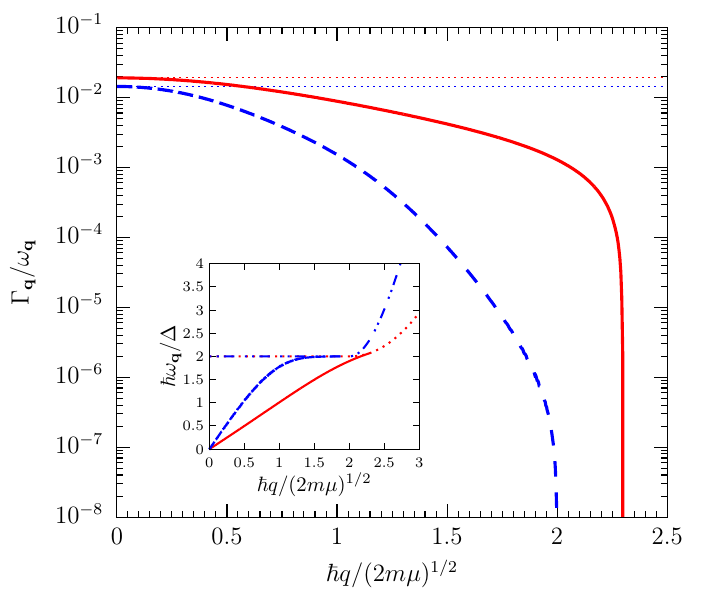}
\end{center}
\caption{\label{fig:GamLU} The inverse quality factor $\Gamma_\qq/\omega_\qq$
of the collective excitations at temperature $\Delta/k_{\rm B}T=5$ 
is shown as a function of $\hbar q/(2m\mu)^{1/2}$ 
at unitarity ($1/a=0$, full red curve) and in the BCS regime ($1/k_{\rm F}a=-0.47$, $\mu/\Delta=2$, blue \resoumis{dashed} curve). The dotted horizontal lines indicate the analytic result 
\eqref{eq:Gammapttsq} in the long wavelength limit, from which the quality factor deviates quadratically at low $q$. 
The damping rate $\Gamma_\qq$
vanishes at the wave number at which the collective branch \resoumis{(full red line ($1/a=0$) and blue dashed line ($1/k_{\rm F}a=-0.47$) of the inset)} hits \cite{CKS2006,CastinCRAS,Concavite} the 
\resoumis{lower edge of the} continuum \resoumis{of fermionic biexcitations} $\kk\to\epsilon_{\kk+\frac{\qq}{2}}+\epsilon_{\kk-\frac{\qq}{2}}$ \resoumis{(dotted red line ($1/a=0$) and blue dash-dotted line ($1/k_{\rm F}a=-0.47$) of the inset)}, that is respectively for $\hbar q/(2m\mu)^{1/2} \simeq2.3$ and $2.0$.}
\end{figure}

\begin{figure}
\begin{center}
\includegraphics[width=0.49\textwidth]{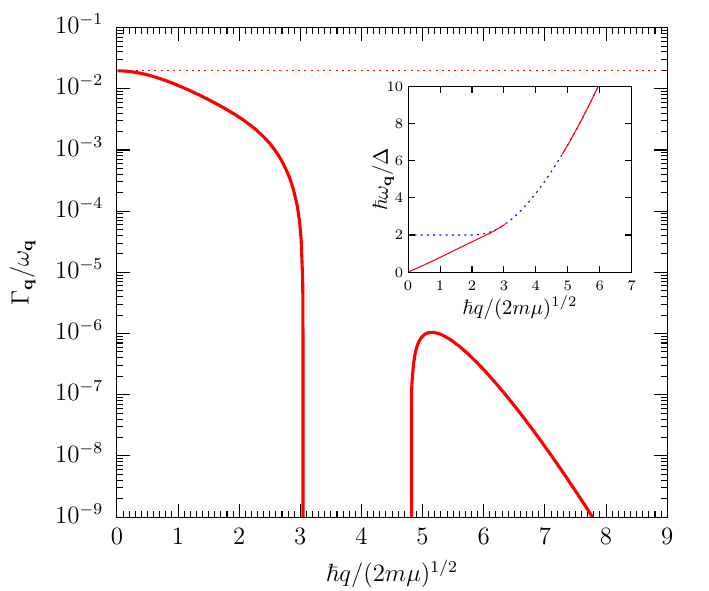}
\end{center}
\caption{\label{fig:GamNonConnexe} 
The inverse quality factor $\Gamma_\qq/\omega_\qq$
of the collective excitations at a temperature $\Delta/k_{\rm B}T=5$ 
is shown as a function of $\hbar q/(2m\mu)^{1/2}$ in the regime
($1/k_{\rm F}a=0.13$, $\mu/\Delta=0.625$) where the
collective branch $q\mapsto\omega_\qq$ (full curve in the inset) is not
simply connected. The branch disappears \cite{CKS2006,Concavite} when it 
reaches the lower edge of the pair-breaking continuum (dotted curve in the inset) at
$\hbar q_{\rm{sup}}/(2m\mu)^{1/2}\simeq3.0$ and reappears as it re-emerges from 
the continuum at $\hbar q_{\rm{inf}}/(2m\mu)^{1/2}\simeq4.8$, after which it
stays close to the continuum edge. The horizontal dotted line shows the 
analytic $q\to0$ limit. The damping rate $\Gamma_\qq$ goes to
zero at $q_{\rm{sup}}$ and $q_{\rm{inf}}$, and remains small in the second
part of the branch ($q>q_{\rm{inf}}$).}
\end{figure}

\resoumis{

\subsection{Towards an experimental observation}

Last, we discuss the experimental observability of the damping rate $\Gamma_\qq$ we predict. Reaching the low temperature regime where our theory is applicable is not anymore a serious limitation: experiments can reach temperatures of a few hundredths \cite{Ketterle2003temperature} of the Fermi temperature $T_{\rm F}$, which, choosing for the gap the value $\Delta/k_{\rm B}T_{\rm F}\simeq0.69$ predicted by BCS theory at unitarity (the value measured by Ref.\cite{Ketterle2008} $\Delta/k_{\rm B}T_{\rm F}\simeq0.44$ is of the same magnitude), corresponds to values of $\Delta/k_{\rm B}T$ of the order of $10$. Similarly, typical experiments have a large enough precision to detect a lifetime of order $1/\Gamma_\qq$ ; for example, choosing $\hbar q/(2m\mu)^{1/2}=0.5$, $1/a=0$ and $\Delta/k_{\rm B}T=5$ (parameters of the full line in Fig.\ref{fig:GamLU}), we have $\hbar\omega_\qq\simeq0.5\Delta$ and $\Gamma_\qq/\omega_\qq\simeq0.015$, which, taking for the Fermi temperature the typical value $T_{\rm F}\approx1\,\mu K$, corresponds to a lifetime $1/\Gamma_\qq\approx 1.5\,\textrm{ms}$. Much longer lifetimes (and quality factors much larger than $\omega_\qq/\Gamma_\qq\simeq67$) have been observed in experiments on the low-energy modes of a cold Bose gas \cite{Dalibard2002}. Similar measurements of the frequency broadening of the collective mode have been performed recently in paired Fermi gas using Bragg excitations  \cite{Vale2017}, but their precision is limited by the spatial inhomogeneity of the gas. We are confident that they will soon reach the precision required to measure $\Gamma_\qq$ thanks to the many ameliorations offered by the flat-bottom potentials \cite{Hadzibabic2013}. 

To conclude on the observability of the damping of the collective excitations by absorption-emission by the fermionic quasiparticles, we still need to compare it to the others three- or four-body damping mechanisms acting on the collective excitations, that are: $(i)$ the Landau-Beliaev processes between three collective excitations \cite{Beliaev1958,Annalen}, $(ii)$ the Landau-Khalatnikov process between four collective excitations \cite{Khalatnikov1949,EPL} and $(iii)$ the process of scattering of a collective excitation on a fermionic quasiparticle \cite{bosonsfermions}. When allowed Beliaev-Landau damping is the dominant phenomenon. For example, keeping $\hbar q/(2m\mu)^{1/2}=0.5$, $1/a=0$ and $\Delta/k_{\rm B}T=5$, we obtain $\Gamma_\qq^{\rm Beliaev-Landau}/\omega_\qq\simeq0.14$; the ratio $\Gamma_\qq/\omega_\qq$ of the process we are interested in is thus one order of magnitude smaller, yet we believe still measurable with a good precision. In fact the situation is even more favorable: the collective branch in superfluid Fermi gases has the particularity of being concave at low wave number on the BCS side \cite{Strinati1998,Concavite} (for $1/k_{\rm F}a<-0.14$ according to RPA) which energetically forbids Beliaev-Landau damping. In this case, only processes $(ii)$ and $(iii)$ remain, which, as far as the damping of collective excitations is concerned, are dominated by the process of absorption-emission by a $\hat{\gamma}$ quasiparticle over a large temperature domain \cite{bosonsfermions}. Taking $1/k_{\rm F}a=-0.47$ (parameters of the dashed line in Fig.\ref{fig:GamLU}, in the near BCS regime where Beliaev-Landau damping is still forbidden) and still $\Delta/k_{\rm B}T=5$ and $\hbar q/(2m\mu)^{1/2}=0.5$, we obtain $8,\!1\times10^{-3}$ for the inverse quality factor of the absorption-emission process, $2.9\times10^{-4}$ for Landau-Khalatnikov process $(ii)$ and $6.6\times10^{-3}$ for the scattering process $(iii)$. The process studied in this article is therefore the main limitation to the lifetime of the collective excitations on the BCS side of BEC-BCS crossover at the usual experimental temperatures. We are thus optimistic about an experimental observation in the near future.}


\section{Conclusion}


The \resoumis{coupling} amplitude for the absorption or emission of a collective excitation by a 
fermionic quasiparticle in a Fermi superfluid is obtained using two complementary
methods, namely the Random Phase Approximation and the functional integral \resoumis{formalism}.
In the limit of long wavelengths our result agrees with the prediction of quantum
hydrodynamics for resonant processes. This result is a necessary ingredient
in order to take interactions between quasiparticles into account in any description
of a Fermi superfluid at low (but non-zero) temperatures. We also calculate
the damping rate of collective excitations, resulting from the coupling to fermionic
excitations beyond the long wavelength limit in the collisionless regime. We show that the inverse quality 
factor decreases (at first quadratically) as the wave number of the collective mode 
increases, and vanishes when the collective mode rejoins the continuum \resoumis{of fermionic biexcitations}. The damping rate we compute can be measured directly using Bragg spectroscopy on the BCS side of the crossover where the collective branch is concave, at the temperatures currently reached by the experiments. 

\acknowledgements 
Discussions with Y. Castin, \resoumis{A. Sinatra}, S. Van Loon and S. Klimin are gratefully acknowledged.
This research was supported by the Bijzonder Onderzoeksfunds (BOF) of the 
University of Antwerp, the Fonds Wettenschappelijk Onderzoek Vlaanderen, 
project G.0429.15.N, and the European Union's Horizon 2020 research and innovation program under the Marie Sk\l odowska-Curie grant agreement number 665501.
\appendix


\section{Integration domain beyond the low-$q$ limit}
\label{ann:domaine}
The main difficulty in evaluating the integral in expression \eqref{eq:Gammaq2}
of the damping rate consists in finding the integration domain $\mathcal{K}$ over the norm of $\kk$. The energy conservation constraint 
$\hbar\omega_\qq +\epsilon_{\kk-\frac{\qq}{2}}-\epsilon_{\kk+\frac{\qq}{2}}=0$ can be rewritten in 
dimensionless form
\be
E(x,u)=0
\label{eq:contrainte}
\ee
with
\begin{multline}
E(x,u)\equiv \sqrt{(x^2+x x_{q} u +x_q^2/4-m)^2 + 1 }  \\ -\sqrt{(x^2-x x_{q} u +x_q^2/4-m)^2 + 1 }  - \hbar\omega_\qq/\Delta
\label{eq:Exu}
\end{multline}
and the dimensionless quantities
\bea
&x^2         =  \frac{\hbar^2 k^2}{2 m \Delta} \label{eq:adim1}
\qquad \qquad &x_q^2                    =  \frac{\hbar^2 q^2}{2 m \Delta} \label{eq:adim1}\\
&m  =  \frac{\mu}{\Delta} \label{eq:adim2} 
\qquad \qquad  &u          = \frac{\kk\cdot\qq}{k q}
\eea
with $x>0$, $m>0$, $x_q>0$ and $u\in[-1,1]$. The function $E$ is taken to depend
explicitly on $x$ and $u$ which represent the integration variables linked to the 
vector $\kk$, but not on $m$ and $x_q$ which are constant parameters fixed by the 
interaction strength and the wave number $q$ respectively. The dimensionless eigenenergy
$\hbar{\omega}_\qq/\Delta>0$ is connected to $x_q$ via the zero-temperature
dispersion relation for the collective mode $q\to\omega_\qq$, obtained by numerically solving 
\eqref{eq:dedispersion}. Our approach consists in finding the value of $u$ that 
satisfies \eqref{eq:contrainte} for fixed $x$,$m$ and $x_q$. This value, if
it exists, is unique since $E$ is a monotonic function of $u$:
\be
\frac{\dd E}{\dd u}=0  \implies  x=0 \textrm{ or } x_q=0 \textrm{ or } m=x^2+x_q^2/4
\ee
For $x=0$ the constraint \eqref{eq:contrainte} can never be satisfied, so this
case can be eliminated. The case $x_q=0$ was discussed in section \ref{sec:petitsq}.
The remaining condition $m=x^2+x_q^2/4$ is independent of $u$ so that 
${\dd E}/{\dd u}$ is either identically zero or has a constant sign
for $u\in[-1,1]$. More precisely,
\be
\forall u\in[-1,1]: 
\begin{cases}
{\dd E}/{\dd u} > 0 \quad \textrm{for} \quad m<x^2+x_q^2/4 \\
{\dd E}/{\dd u} <0 \quad \textrm{for} \quad m>x^2+x_q^2/4
\end{cases}
\ee
Moreover, since we have $E(x,0)=\hbar\omega_{\textbf{q}}/\Delta,\ \forall x>0$, 
the function $u\mapsto E(x,u)$ has a root in $[-1,1]$ if and only if
\be
\begin{cases}
E(x,-1) < 0 \quad \textrm{for} \quad  x>\sqrt{m-x_q^2/4} \label{eq:conditonx1}\\
E(x,1) < 0  \quad \textrm{for} \quad x<\sqrt{m-x_q^2/4}
\end{cases}
\ee
These conditions determine the set $\mathcal{K}$. Close to $x=\sqrt{m-x_q^2/4}$
neither condition is met, because in this point $E(x,u)$ is constant and strictly positive.
This implies that the integration domain splits up as in the low $q$ limit
into $\mathcal{K}_1\subset[\sqrt{m-x_q^2/4},+\infty[$ and 
$\mathcal{K}_2\subset[0,\sqrt{m-x_q^2/4}]$. 
Since in the limit $x\to\infty$, 
$E(x,-1)$ goes monotically to $-\infty$ for $x>\sqrt{m-x_q^2/4}$, the
upper domain $\mathcal{K}_1$ is of the form $[x_1,\infty[$ with a lower bound $x_1$ that
can be determined by numerically solving $E(x,-1)=0$ for
$x\in[\sqrt{m-x_q^2/4},+\infty[$. 
Since $E(0,1)>0$ and $E(\sqrt{m-x_q^2/4},1)>0$, the domain $\mathcal{K}_2$ is of the form $[x_2,x_3]$ with $0<x_2<x_3<\sqrt{m-x_q^2/4}$. Contrarily to $\mathcal{K}_1$, it does not exist for all values of $m$ and $x_q$ (it clearly disappears for $m-x_q^2/4<0$). To know whether it exists, we look for the minimum $x_{\rm min}$ of $E(x,1)$ in $[0,\sqrt{m-x_q^2/4}]$. If at that point the function is of negative sign $E(x_{\rm min},1)<0$, we look for the boundaries of $\mathcal{K}_2$ such that $x_2<x_{\rm min}<x_3$. Otherwise $\mathcal{K}_2$ is the empty set.

\bibliography{/Users/rox/Documents/biblio}
\bibliographystyle{unsrtnat}
 \end{document}